# Analysis of Group Multiuser Detection Based on Coalitional Game

Cuilian Li[1,2], Zhen Yang[1]
(1. Institute of Signal Processing and Transmission, Nanjing University of Posts and Telecommunications, Nanjing, Jiangsu 210003, China; 2. Zhejiang Wanli University, Ningbo, Zhejiang 315100, China)

**Abstract**[1]
The issue of group-blind multiuser detection in MAC channel among wireless nodes in the environment of multiple networks coexisting and sharing spectrum is addressed under the Framework of coalitional game. We investigate the performance and stability of multiple access channel (MAC) with linear decorrelating multiuser detection under varying SNR, channel gains and coalitional structures, in which both single BS and multiple BSs cases were considered. The main results and conclusion are as follows: (1) the grand coalition is payoff maximizing under loose SNR; (2) it is in conformity with group and coalitional rationality forming coalition among nodes that have comparative channel gains.

## 1 Introduction

The dynamic sharing of spectrum has been at the forefront of radio regulation and efficient system design since the earliest days of wireless. Given that devices sharing the spectrum may differ vastly in their modulation schemes and protocols, especially in the unlicensed spectrum bands, receiver cooperation may only be feasible if one could find ways to jointly process their signals. A spectrum server (SS, such as the one proposed in [2] can serve as a central entity assembling and processing data, that enables disparate devices to jointly decode their signals, avoid interference and share spectrum resource more efficiently.

There are abundant network and communication researches based on game theory. Yet to the best of our knowledge, in wireless network domain, few published results of cooperative game, especially coalitional game research can be found. In [9] Z. Han et al. proposed an approach based on coalition Games to overcome the problem known as the curse of the boundary nodes, and in [8] they proposed a fair scheme to allocate subcarrier, rate, and power for multiuser orthogonal frequency-division multiple-access systems. The approach considers a new fairness criterion, which is a generalized proportional fairness based on Nash bargaining Solutions and coalitions. N. Xia et al. propose a WSN coalition formation algorithm based on Ant Colony System with dual-negative feedback characteristic in [5]. They noted that their algorithm can balance the energy consumption among sensor nodes, so as to extend the network lifetime for more tasks. S. Mathur et al. studied the issue of sharing spectrum through receiver cooperation in wireless networks under the framework of coalitional game theory in [6, 7].

In this paper, we analyze the effect of cooperative group MUD under MAC channel based on the analytical framework of coalitional game theory, we compare the gains/payoff and stability of variety coalitional structures in view of channel condition, SNR etc. in particular, we address the above questions take into consider both simplified single BS network models and the more complex multiple BSs scene.

## 2 Coalitional game (CG)

Coalitions are formed among players in view of their individual rationality and group rationality. Coalitional Games come into two main guises, depending on whether the worth of a coalition can be freely distributed amongst its members or not: Games with transferable payoffs (TU-Games), where the worths are transferable amongst players forming a coalition without any limitation, and Games with non-transferable payoffs (NTU-Games). The MUD coalitional game model here belongs to NTU-Game and thus we only present the definition of NTU-Games:

*Definition 1.* A Coalitional Game without transferable payoff is a four-tuple $< N, X, v, (\geq_i)_{i \in N} >$, where:

➢ N is a finite set of players;
➢ F is the set of all possible consequences;
➢ $v : s \to 2^X$ is a function that assigns, to any coalition $s \subseteq N$ of players, a set of consequences $v(S) \subseteq X$;
➢ $(\geq_i)_{i \in N}$ is the set of all preference relations $\geq_i$ on X, $\forall i \in N$.

Core is probably the most important solution concept defined for such games, and here only definitions for NTU-games are provided next:

*Definition 2.* The core of the coalitional game without transferable payoffs $< N, X, v, (\geq_i)_{i \in N} >$ is the set of all $\bar{x} \in v(N)$ such that there is no coalition $s \subseteq N$ with a $\bar{y} \in v(s)$ such that $\bar{y} \succ_i \bar{x}$ for all $i \in s$.

## 3 Group-blind multiuser detection (GMUD) [3] under the framework of coalitional game theory

We compare the performance and stability of variety coalitional structures under the solution concept of core in view of channel condition, SNR etc.

### 3.1 Signal/Network model

Consider a synchronous multi-cell code-division multiple-access (CDMA) system with a multiuser receiver, that uses binary phase-shift keying (BPSK) modulation, assuming B base stations with omni-directional antennas. All users are assumed to have symmetric signature sequences, i.e., the cross correlation between any two signature waveforms from different users are the same: $\rho_{ij} = \rho$, $0 \leq \rho < 1$, for all $i \neq j$. Note the transmission power of each user as P. For each BS receiver, there are two kinds of users.

---

[1] Supported by National Natural Science Foundation of China (No.60772062), the Key Projects for Science and Technology of MOE (No.206055) and the Key Basic Research Projects for the Natural Science of Jiangsu Colleges (No.06KJA51001).

> Known users: For these, the receiver knows their signature sequences, and uses them in the detector structure. The system can control their transmission power.
> Unknown users: For these, the receiver does not know their signature sequences, and cannot control their transmission power. The system does not detect their signals.

Assume there are totally K users in the network, of which M are known and N=K－M are unknown for each BS. For now we consider only the cooperative form of GMUD in BS. Each BS perform joint detection for the known users using their signature sequences and estimated channel gains, since eliminate interferences from other MSs in the MUD group and improve signal-to-noise-interference ratio (SINR) of the received signals. In this paper only linear decorrelation multiuser detection [3] is considered.

### 3.2 Decorrelator GMUD coalitional game

For MAC channel, Multiuser receivers suppress the interference between users in spread-spectrum CDMA systems by making use of the structure of the multiple-access interference and of the knowledge of the code sequences, thus achieve a gain in SINR. As the gain of each user in SINR by multiuser detection can't possibly share among MSs, owing to the technical characteristic of MUD, this cooperation can only model as NTU-Game [4].

Taking the SINR of received signal as the respective MS's payoff, then for the network model in section 3.1, the M MSs under each BS can be considered as an coalitional game $<M, SINR, R^M, (\geq_i)_{i \in M}>$, in which M is the set of players/MSs, SINR is the allocation function, $R^M$ is the result set, a real vector of M dimension and defining preference $(\geq_i)_{i \in M}$ as ordinal relation on real domain.

The SINR (payoff allocation) of a MS when cooperates in the GMUD game can represent as [3]:

$$SINR_i^{decorr}(\Omega) = \frac{P_i}{\frac{\sigma^2}{1-\rho}\frac{1+\rho(|\Omega|-2)}{1+\rho(|\Omega|-1)} + [\frac{\rho}{1+\rho(|\Omega|-1)}]^2 \sum_{i \in \Omega^c} P_i},$$

$\forall i \in \Omega$, (1)

in which $\Omega$ is the formed coalition (the set of MSs which cooperate), and $\Omega^c$ is the set of MSs outside of the coalition. $P_i = h_i^2 P$ is the received power of the signal of MS i at BS, $h_i$ is the channel gain of MS i, and $\sigma^2$ is the variance of the zero mean white Gaussian noise.

The SINR of a MS when doesn't cooperate (that is, when BS uses matched filter receiver) is:

$$SINR_i^{mf} = \frac{P_i}{\sum_{j \neq i, j \in \Omega} \rho^2 P_j + \sum_{j \in \Omega^c} \rho^2 P_j + \sigma^2} \quad (2)$$

Among the basic assumptions of cooperation game are coalitional rationality and group rationality, in addition to individual rationality. As a player, a MS's individual rationality is to maximize its SINR. Here we take the sum SINR of all MSs in the coalition as the overall payoff as the group, then the coalitional rationality and group rationality of a MS is to seek the maximization of the sum SINR of a group or the grand coalition. Stable coalitions are those coalitional structures that meet all the above three rationality at the same time.

**Fading model**

The considered path loss model includes both distance based path loss and shadowing.

$$|h_i|^2 (dB) = \frac{P_r}{P_t}(dB) = K - 10\mu \log_{10} d - \psi_s \text{ (dB)} \quad (3)$$

in which $h_i$ denotes the dB path loss of the upward channel of MS i, $K$, $\mu$ is the path loss constant and exponent respectively, and $\psi_s$ is the shadowing, a zero mean, $\sigma_s$ variance Gaussian random variable.

For coalitional game, the number of all possible coalition structures is a function of m = |M|, the members of the game, and can be obtained from the Bell numbers. The Bell number Bn is equal to the number of ways a set of n elements can be partitioned into non-empty subsets. As the number of possible partitions increases extremely quickly with an increase of m, we only consider the cases of a few MSs, more complicated cases can be analyzed analogously.

### 3.3 Case of single BS

Considering the case that there are m=3 MSs under the BS, $B_3$=5 and possible coalitional structures in this case are: {123，12|3，1|23，13|2，1|2|3}.

Fig.1 to fig.4 are simulation results of coalition payoff of each coalition structure under the influence of parameters such as SINR, shadowing and path loss exponent (MATLAB).

Fig.1 is the positions of MSs relative to BS. For the fading model given in (3), the distance of each MS to BS can approximately represent their signals strength received at BS. Bars in fig.1(b) give the mean value of 10 simulations, to mitigate the influence of channel randomness on the payoffs. It shows stochastically that:

> Comparing with other coalition structures, the grand coalition 123 is payoff maximizing both from the point of individual rationality and group rationality. So the only stable structure is the grand coalition and thus the core of this game is {v([123])}.

➢ All the coalitional structures are not dominated in the sense of individual and group payoffs by the non-cooperative case 1|2|3.

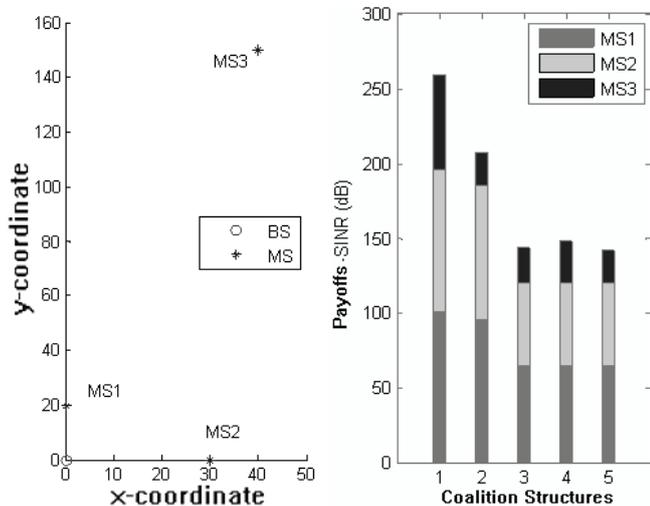

Fig.1 Single BS case **(a)** Layout of BS and MSs **(b)** Payoff of each MS under each coalition structure when SNR=27dB(the 1、2、3、4 and 5 in x-coordinate denote coalition structure 123、12|3、1|23、13|2 and 1|2|3 respectively.)

Fig.2 compares the group payoffs of GMUD game of each coalitional structure when SNR (the ratio of MS's transmitting power with noise power) increases from -40dB to 40dB. 3(b) shows gains of each structure over non-cooperative case. It's evident that:

➢ No cooperation structures are dominated by non-cooperation case under all SNR. When SNR>-20dB, collaborating has notable dominance over non-cooperating from the point of group rationality.

➢ Payoff of coalition structure 12|3 is very close to that of the grand coalition and has distinct dominance over other structures, while the payoff of structure 1|23 nearly overlaps with that of non-cooperating.

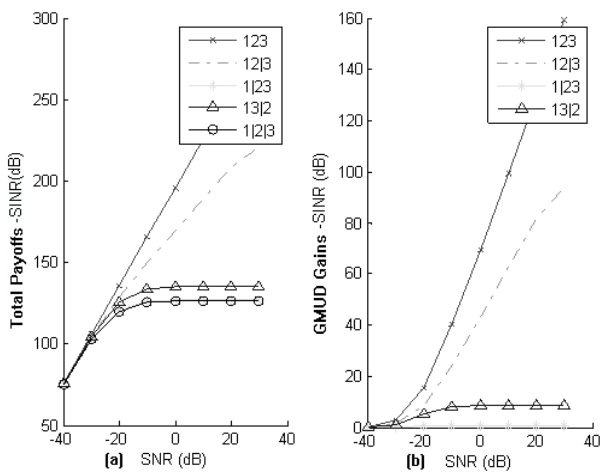

Fig.2 Single BS case **(a)** Total payoffs of each coalition structure with respect to SNR **(b)** Total payoff gains of each coalition structure over non-cooperative case (1|2|3) with respect to SNR

Fig.3 shows the group payoffs of GMUD game of each coalitional structure with respect to shadowing. It can be seen that the more severe the shadowing, the more fluctuating the payoffs, yet their relative positions are still same as those in fig.1 and fig.2. Fig 4 shows the effects of path loss exponent $\mu$ on the total payoff of each coalitional structure. The distribution of $\mu$ depends on distinct radio environments[1]. It is about 2 in free space, about 3 in urban cells or indoor. It can be seen that when $\mu$ >6, there is hardly any advantage to cooperate, when $\mu$ is very small (near 0), the grand coalition has prominent advantage over otherwise and in typical urban environment, both structure 12|3 and the grand coalition are dominating.

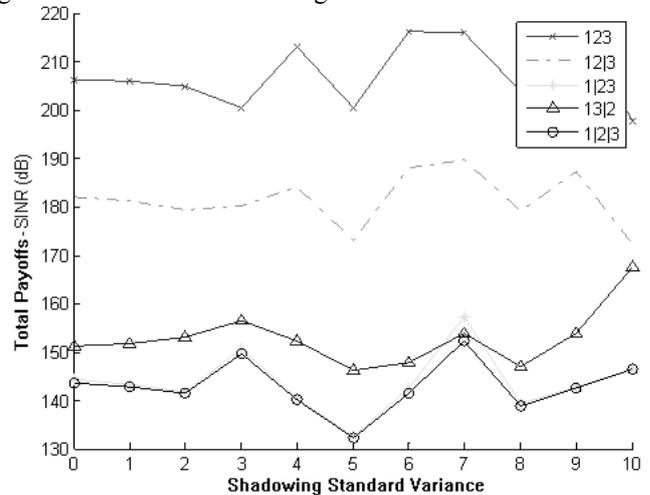

Fig.3 Total payoffs of each coalition structure with respect to shadowing

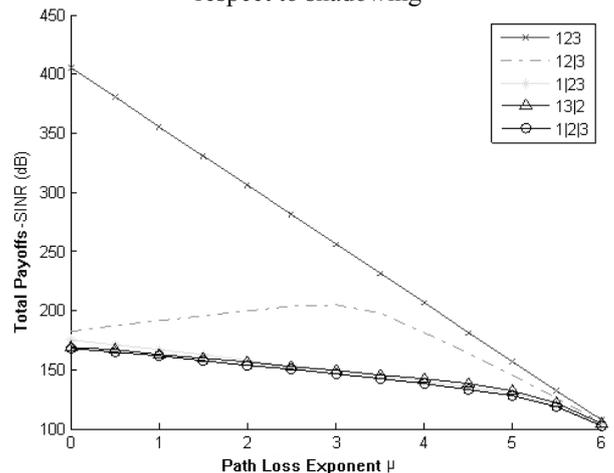

Fig.4 Total payoffs of each coalition structure with respect to path loss exponent

### 3.4 Case of multiple BSs

We consider in this section whether the research results can be generalized to system with multiple BSs. We don't take into consider the inter-BSs cooperation for now. For a specific BS, the MSs under other BS are unknown users; the signals from these MSs are regarded as interference.

Considering the case of 2 BSs and there are 4 active MSs under each of them. Then for each BS, the possible number of coalitions is B4=15, namely, {1234，123|4，124|3，134|2，1|234，12|3|4，13|2|4，14|2|3，23|1|4，24|1|3，1|2|34，12|34，13|24，14|23，1|2|3|4}. For simplicity, based on the conclusion of last section and the layout of BSs and MSs in our simulation, we merely select and analyse some representative coalition structures.

We give the simulation results of 2 BSs case in fig.5. Fig.5 (a) shows the network topology, in which each BS has 2 near MSs and two far nodes respectively. For this network layout, in view of our results in last section, we focus our attention on the following coalition structures: the grand coalition 1234, coalition structure 12|3|4 (indicating the case that the two near nodes cooperate and the two far nodes don't), structure 2|34 (the two near MSs and the two far MSs form two non-overlapping coalition respectively) and the non-cooperating case 1|2|3|4. Fig.5(b),(c) show total payoffs of coalition model BS1 and BS2 under the above four coalition structures when SNR (same as section 3.3) increases from -60dB to 20dB. It's obvious that:

➢ No cooperation structures are dominated by non-cooperation case under all SNR---same as the single BS case in section 3.3.
➢ And the grand coalition case has the best payoff comparing with other coalition structure.
➢ The total payoff of 12|3|4 is nearly overlapping entirely with that of 12|34, indicating that it is of little influence to the total payoff whether the far nodes cooperate or not.

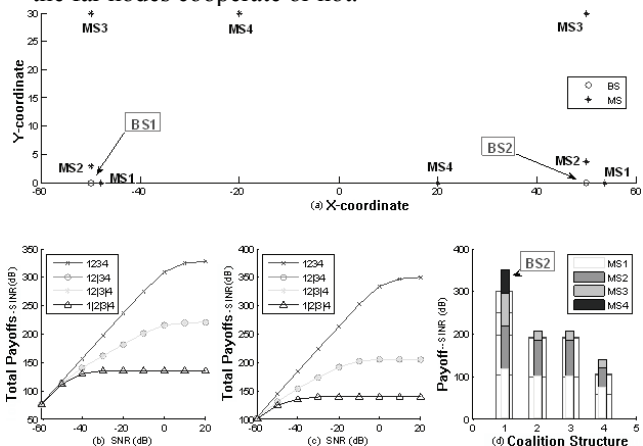

Fig.5 Two BSs, each with 4 MSs case
**(a)** Layout of BSs and MSs **(b)** BS1: Total payoffs of each coalition structure with respect to SNR **(c)** BS2: Total payoffs of each coalition structure with respect to SNR **(d)** Payoff of each MS under each coalition structure when SNR=20dB for both BS1 and BS2 (the 1、2、3 and 4 in x-coordinate denote coalition structure 1234、12|3|4、12|34 and 1|2|3|4 respectively.)

Fig.5(d) shows the total payoffs of both BSs as well as the payoff of each MS under four coalition structures when SNR=20dB. As the dB value of SINR can be negative, to facilitate comparing in histogram, we add |SINRmin| (the absolute value of the minimum SINR of all MSs) to all SINRs. Thus if SINR of some MS is absent from the figure, it means that the value of it is SINRmin. It can be seen that:

➢ Comparing with other coalition structures, the grand coalition 1234 is payoff maximizing both from the point of individual rationality and group rationality.
➢ All the coalitional structures are not dominated in the sense of individual and group payoffs by the non-cooperative case 1|2|3, while the gains of near nodes are more distinct.
➢ There are little differences between the payoffs of 12|3|4 (both total payoff and individual payoff) and those of 12|34, indicating that it is of little influence to total and individual payoff whether the far nodes cooperate or not.

## 4 Conclusions

We investigate MAC channel group multiuser detection payoffs under the coalitional game theory framework, considering both single BS and two BSs cases. Specifically we compare the total and individual payoffs of various coalition structures in view of different SNR and channel conditions. We acquire important results as to the payoffs and stability of each coalition structure: 1. Cooperation in GMUD form can be profitable in view of both group and individual rationality; 2. It is more effective for MSs with similar channel condition to cooperate when taking both cooperative payoff and cost into account. Further research may include cooperation among BSs, whether GMUD cooperation can be modelled as TU-Game and if so, when.

## References
[1] A. Goldsmit. Wireless Communication, Cambridge University Press, (2005)
[2] C. Raman,R.D. Yates, et al. "Scheduling variable rate links via a spectrum server", *DySPAN 2005*.: 110-8, (2005)
[3] L. Yun, A. Ephremides. "Linear multiuser detectors for incompletely known symmetric signals in CDMA systems." *IEEE Transactions on Information Theory* 50 (6): 1026-40, {2004).
[4] M. Osborne, A. Rubenstein. A Course in Game Theory, MIT Press, (1994)
[5] N. Xia, J. Jiang, and et al. "A WSN Coalition Formation Algorithm Based on Ant Colony with Dual-Negative Feedback". http://dx.doi.org/10.1007/978-3-540-72590-9_170, (2007)
[6] S. Mathur, S. Lalitha, et al. "Coalitional Games in Receiver Cooperation for Spectrum Sharing", *2006 40th Annual Conference on Information Sciences and Systems*: 949-4, (2006)
[7] S. Mathur, L. Sankaranarayanan, et al. "Coalitional Games in Cooperative Radio Networks", *ACSSC '06*.: 1927-1, (2006)
[8] Z. Han, Z. Ji, et al. "Fair Multiuser Channel Allocation for OFDMA Networks Using Nash Bargaining Solutions and Coalitions." *IEEE Transactions on Communications* 53 (8): 1366-6, {2005).
[9] Z. Han, H.V. Poor. "Coalition Games with Cooperative Transmission: A Cure for the Curse of Boundary Nodes in Selfish Packet-Forwarding Wireless Networks". http://arxiv.org/abs/0704.3292,(2007)